\documentclass{PoS}
\usepackage{amsmath}
\usepackage{graphicx}
\usepackage[square,numbers]{natbib}
\usepackage{subfig}

\usepackage{epsfig} 
\usepackage{ifpdf} 
\usepackage{fancyvrb}
\usepackage{txfonts}
\usepackage{xcolor}
\usepackage{soul}
\usepackage{mathtools}

\title{5.5 years multi-wavelength variability of Mrk 421: evidences of leptonic emission from the radio to TeV}

\ShortTitle{5.5 years multi-wavelength variability of Mrk\,421}

\author{
\speaker{V.\,Sliusar}$\,^1$,  A.\,Arbet-Engels$\,^2$,  D.\,Baack$\,^3$,  
M.\,Balbo$\,^1$,
M.\,Beck$\,^{2,a}$,  N.\,Biederbeck$\,^3$,
A.\,Biland$\,^2$,   M.\,Blank$\,^4$,  
T.\,Bretz$\,^{2,a}$,
K.\,Bruegge$\,^3$,  M.\,Bulinski$\,^3$,  J.\,Buss$\,^3$,  
M.\,Doerr$\,^4$,
D.\,Dorner$\,^4$,  D.\,Elsaesser$\,^3$,  D.\,Hildebrand$\,^2$,  
R.\,Iotov$\,^4$,
M.\,Klinger$\,^{2,a}$, K.\,Mannheim$\,^4$, S.A.\,Mueller$\,^2$, 
D.\,Neise$\,^2$,
M.\,Noethe$\,^3$,  A.\,Paravac$\,^4$,  
W.\,Rhode$\,^3$,
B.\,Schleicher$\,^4$,  K.\,Sedlaczek$\,^3$,  A.\,Shukla$\,^4$,  
L.\,Tani$\,^2$,
F.\,Theissen$\,^{2,a}$ and
R.\,Walter$\,^1$
(the~FACT~Collaboration)\\

    {$^1$}University of Geneva, Department of Astronomy,
     Chemin d'Ecogia 16,  1290 Versoix, Switzerland\\
    {$^2$}ETH Zurich, Institute for Particle Physics and Astrophysics,
     Otto-Stern-Weg 5, 8093 Zurich, Switzerland\\
    {$^3$}TU Dortmund, Experimental Physics 5,
     Otto-Hahn-Str. 4, 44221 Dortmund, Germany\\
    {$^4$}Universit\"at W\"urzburg, Institute for Theoretical Physics and 
Astrophysics,
     Emil-Fischer-Str. 31, 97074 W\"urzburg, Germany\\
    {$^a$}also at RWTH Aachen University, Physics Institute III A, 52074 
Aachen, Germany\\
   E-mails: \email{vitalii.sliusar@unige.ch}, 
\email{roland.walter@unige.ch} \medskip\\
}

\abstract{
Mrk 421 is a high-synchrotron-peaked blazar featuring bright and persistent GeV and TeV emission. We use the longest and densest ongoing unbiased observing campaign obtained at TeV and GeV energies during 5.5 years with the FACT telescope and the \textit{Fermi}-LAT detector. The contemporaneous multi-wavelength observations were used to characterize the variability of the source and to constrain the underlying physical mechanisms. We study and correlate light curves obtained by nine different instruments from radio to gamma rays and found two significant results. The TeV and X-ray light curves are very well correlated with lag, if any, shorter than a day. The GeV light curve varies independently and accurately leads the variations observed at long wavelengths, in particular in the radio band. We find that the observations match the predictions of leptonic models and suggest that the physical conditions vary along the jet, when the emitting region moves outwards.
}

\FullConference{High Energy Phenomena in Relativistic Outflows VII (HEPRO VII)  -HEPRO2019-\\
		July 9th - July 12th, 2019\\
		Barcelona, Spain}

\begin{document}

 \makeatletter
 \setbox\@firstaubox\hbox{\small V.~Sliusar}
 \makeatother

\section{Introduction\label{sec:introduction}}

Mrk\,421 is a bright nearby high-energy-peaked blazar ($z = 0.031$), perfectly suitable for the study of the broadband emission from a relativistic jet. Mrk\,421 features frequent and bright GeV and TeV flares. Its spectral energy distribution (SED) has two humps peaking in the X-rays and in GeV energies. The precise model of the blazar emission is still under discussions. It is generally agreed that the low energy hump is created by relativistic emitting electrons, though the high energy one has been explained by hadronic or leptonic emission, or by a mixture of the two. A one-zone leptonic synchrotron self-Compton model (SSC) \citep{abdo_2011ApJ...736..131A} can be used to explain the complete SED of Mrk\,421. Hadronic emission has been proposed as a result of proton synchrotron \citep{2015MNRAS.448..910C}, hadrons can also interact with the leptonic synchrotron photons creating cascades of pions and muons, which then decay and emit $\gamma$-rays and neutrinos \citep{2001APh....15..121M}. 

The X-ray and TeV variability of Mrk\,421 suggests either that relativistic shocks move within the jet or that the emission region is much smaller than the gravitational radius \citep{2014A&A...563A..91A}, possibly driven by the interactions between stars/clouds and the jet, or by magnetic reconnections.

Mrk\,421 has been the target of numerous multi-wavelength (MWL) campaigns aiming to investigate the emission processes within its jet \citep[e.g.][]{2004ApJ...601..759T,2015A&A...576A.126A}. Most studies reported strong correlation between the X-ray and TeV emission, with a maximum lag of $\sim$5 days \citep{2015A&A...576A.126A}. The highest variability was found in X-rays, where also a harder-when-brighter behaviour was discovered.

Here, we report results of variability and correlation studies, based on about 5.5 years long multi-wavelength monitoring campaign, which includes dense TeV observations with the First G-APD Cherenkov Telescope (FACT) \citep{2013arXiv1311.0478D}. Unlike other TeV observational campaigns, FACT was not triggered in case of flare detected elsewhere, but observed densely and regularly over the years. At lower energies, we used continuous radio, optical, ultraviolet, X-ray, and GeV data, obtained quasi-simultaneously with FACT observations.

\section{Multi-wavelength data\label{sec:data}}

Nine different instruments contributed data to the MWL dataset used in this study. The light curves span from December 14, 2012 to April 18, 2018. During this time period, Mrk\,421 was observed in various flux states. Flares and increased flux were observed in all considered bands. Some flares, observed in the X-ray and TeV band, lasted for a few days, so could be individually identified (see table~\ref{tab:flares}). 

FACT is a 3.8\,m imaging air Cherenkov telescope located in La Palma, Spain \citep{2013JInst...8P6008A}, operating since late 2011 \citep{2013arXiv1311.0478D}, and remotely controlled since July 2012 (fully robotic observations were achieved in late 2017). Its SiPM camera and the state-of-the-art feedback system\citep{2014JInst...9P0012B}, allows to observe even in bright ambient light conditions \citep{2013arXiv1307.6116K}. Analysis techniques and quality checks, which were applied to the Mrk\,421 data, are reported in \citep{2017ICRC...35..779H, 2017ICRC...35..612M, 2019arXiv190203875B, 2005ICRC....5..215R}. Using simulated data, the energy threshold of the telescope for this source is estimated as 850\,GeV.

We used data from the Fermi Large Area Telescope (LAT) \citep{2009ApJ...697.1071A} to cover the GeV $\gamma$-rays band with energy 100 MeV $<$ E $<$ 300\,GeV. The PASS8 pipeline, Fermi Science Tool v10r0p5 package, and background sources from the LAT 4-year Point Source Catalogue were used for the modelling.

X-ray data originates from multiple space-based telescopes instruments: Swift/BAT, Swift/XRT and MAXI. Swift/BAT covers 15-50\,keV band, data reduction pipeline is based on the BAT analysis software \texttt{HEASOFT} version 6.13 \cite{2013ApJS..207...19B}. Using Swift/XRT \citep{2005SSRv..120..165B} data, we obtained the light curves for the 0.3-2\,keV and 2-10\,keV bands directly from the on-line Swift-XRT products generation tool \footnote{http://www.swift.ac.uk/user\_objects/} which adopts the \texttt{HEASOFT} software version 6.22. MAXI \citep{2009PASJ...61..999M} is sensitive from 2\,keV to 20\,keV, and the light curve for Mrk\,421 is publicly available \footnote{http://maxi.riken.jp/star\_data/J1104+382/J1104+382.html}.

Optical observations were performed by multiple ground- and space-based telescopes. In the V-band, as a part of a blazar monitoring campaign \citep{2009arXiv0912.3621S}, Mrk\,421 was observed by the 1.54\,m Kuiper and Bok telescopes. Data from Cycle 5 to 10 are available online\footnote{http://james.as.arizona.edu/$\sim$psmith/Fermi/DATA/photdata.html} and was used in this study. In the UV, the source is observed by Swift/UVOT \citep{roming_2005SSRv..120...95R}. The on-, off-method using the HEASOFT package version 6.24 along with UVOT CALDB version 20170922 was adopted for the data reduction.

Regular observations of Mrk\,421 at 15\,GHz were performed by the the OVRO 40 meter radio telescope. The light curve is available through the public archive\footnote{http://www.astro.caltech.edu/ovroblazars/}.

\section{Timing and correlation analysis}

We performed  cross-correlation, auto-correlation, Bayesian Block and fractional variability analysis of the considered MWL light curves to investigate the physical processes responsible for the emission in all these bands. We also estimated the best-fit parameters of the GeV to radio response profile to generate a synthetic radio light curve from the GeV one.

\subsection{Fractional variability}
Following the prescription in \citep{Vaughan_2003MNRAS.345.1271V} and uncertainties estimation suggested in \cite{Poutanen_2008MNRAS.389.1427P} the fractional variability analysis was performed for all Mrk\,421 light curves. The highest variability ($F_{var}=1.33$) was found in the hard X-rays (Swift/BAT) and the lowest in the radio ($F_{var} = 0.15$). Beyond the X-ray band, the variability drops at GeV energies ($F_{var}=0.34$) and increases again at TeV energies ($F_{var}=0.92$). The two distinctive humps in the variability spectrum match the SED, as already previously reported \citep{2015A&A...576A.126A, aleksic_2015A&A...578A..22A}. The two emission humps are more variable towards higher energies. As simultaneous (within a day) X-ray and TeV fluxes are correlated, a single parameter (the cutoff energy) is indeed driving the main variations observed in both components.

\subsection{Light curves correlations}
Long term unevenly sampled light curves allow to estimate correlations for sub-sampling time lags. We performed such an analysis using the discrete cross-correlation function (DCF) as proposed in \citep{Peterson_1998PASP..110..660P}. Originally underestimated uncertainties of the DCF were recalculated using flux randomisation and random subset selection process for MC simulations. The final lag between the light curves is estimated using the centroid method \citep{Peterson_1998PASP..110..660P, 2003ApJ...584L..53U} at 80\% maximum. The variability time scale observed in the TeV and X-ray bands are short, $\sim$3 days, which is consistent with the models where the emission in the jet is dominated by relativistic electron cooling.

\begin{figure}
  \centering
       \includegraphics[width=0.83\columnwidth]{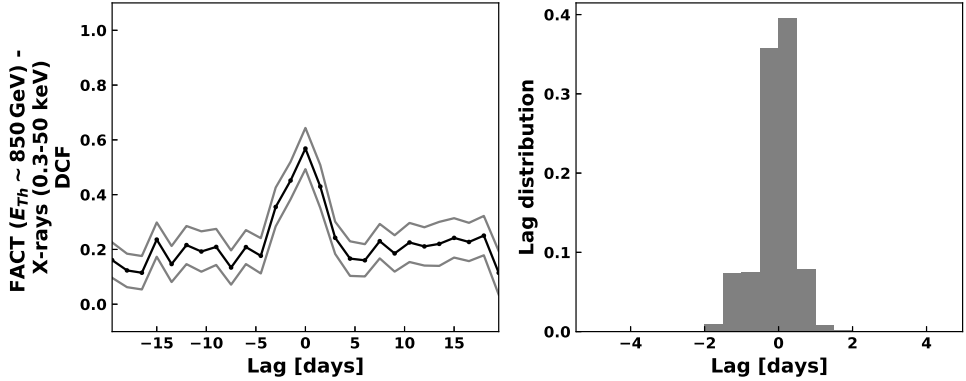}
       \caption{Combined DCF cross-correlations of TeV (FACT) and X-rays (MAXI, Swift/BAT, Swift/XRT) variability. Left: DCF values as a function of lag. Gray lines denote the 1$\sigma$ uncertainties. Right: lag distribution corresponding to the maximum DCF value.}
  \label{fig:dcf_all}
\end{figure}

A strong cross-correlation was found between the TeV and X-ray light curves. The combined (all X-ray bands) X-rays lag is ($0.02\pm0.41$) days ($1\sigma$). The combined cross-correlation between X-ray  and TeV bands and respective lag distribution are shown in Fig.~\ref{fig:dcf_all}. This result is more constraining than previously reported lags on more sparse and shorter data sets \citep{2015A&A...576A.126A}. The X-ray and TeV light curves are not correlated with the GeV, optical or radio light curves. The low energy bands are correlated with the GeV band although with wide response and lags of 40 to 70 days.

A Bayesian block algorithm was applied to the X-ray and TeV light curves to identify individual flares assuming simultaneous flares in these two bands. A flare was defined as an increased flux for at least two days with amplitude at least twice the amplitude of the previous Bayesian block. The identified flares are listed in table \ref{tab:flares}. 29 of the TeV flares were coincident with ones in X-rays. This indicates that particle populations with different energy distributions are necessary to explain all the observations.

\begin{table}[h!]
\centering
\caption{List of TeV and X-ray flares sorted by the spectral bands in which they were detected.}
\label{tab:flares}
\begin{tabular}[t]{lcp{7cm}}
Bands & Number & Time ranges, MJD\\
\hline
\hline
TeV only: & 2 & 56689-56692, 57006-57015  \\ \hline
TeV and X-rays: & 29 & 56317-56330, 56369-56383, 56389-56400, 56441-56449, 56650-56670, 56696-56700, 56751-56755, 56976-57005, 57039-57053, 57065-57070, 57091-57099, 57110-57121, 57188-57192, 57368-57379, 57385-57390, 57422-57431, 57432-57449, 57504-57511, 57531-57538, 57545-57550, 57728-57753, 57756-57769, 57770-57775, 57787-57793, 57850-57866, 58103-58113, 58129-58142, 58162-58167, 58185-58196  \\ \hline
\end{tabular}
\end{table}

\subsection{GeV to radio response}

\begin{figure}
  \centering
      \includegraphics[width=0.83\columnwidth]{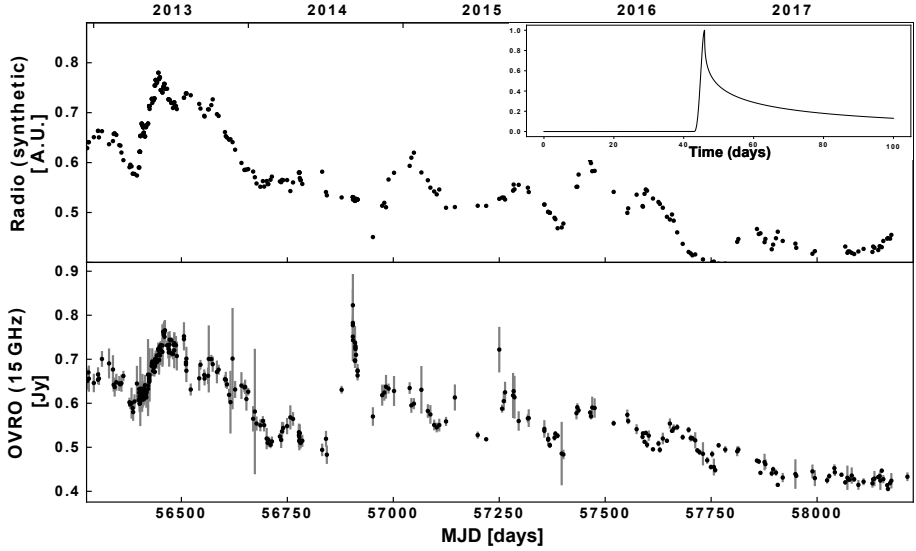}
  \caption{Synthetic radio light curve (top) derived from the \textit{Fermi}-LAT light curve and OVRO 15\,GHz radio light curve (bottom). The GeV to radio response profile is depicted in the top right corner. The $y$ axis is in arbitrary units.}
  \label{fig:fermi_radio_conv}
\end{figure}

Since there is a wide correlation between the GeV and radio light curves, we attempted to reproduce the radio light curve by convolving the GeV light curve with a specially constructed response profile \cite{turler_1999A&A...349...45T}. The response profile, original and synthetic radio light curves are shown in Fig.~\ref{fig:fermi_radio_conv}. We found that the response has $t_{rise}=3$ days, $t_{fall}=7.7$ days, for a delay of $\Delta t = 43$ days. Need for such a delay was also reported previously in 3C\,273 \citep{esposito_2015A&A...576A.122E}. Adopting such an approach, we were able to well reproduce the radio light curve except for one very fast radio flare (MJD 56897).

\section{Results and conclusions\label{sec:conclusions}}

We performed correlations, variability and timing analysis of 5.5 years of dense multi-wavelength data of Mrk\,421 and found that:

\begin{enumerate}
\item The highest variability is exhibited in the X-rays and at TeV energies. 95\% of the short X-ray and TeV flares are coincident. The lag between the TeV and X-ray light curves is not significant and is shorter than ($0.02\pm0.41$) days.

\item The radio light curve can be reproduced by convolving the GeV light curve with a delayed fast rising and slowly decaying response profile.

\end{enumerate}

The fractional variability of Mrk\,421 and the correlated TeV and X-ray emission indicate that the main source of variability is dominated by a synchronous change of the cutoff energies of the low and high-energy components. The zero lag correlation between the TeV and X-ray light curves of Mrk\,421 indicates that these two emissions are driven by the same physical parameter and are consistent with a leptonic emission scenario. The observed X-ray and TeV variability time scales do not match the prediction for proton cooling for shock acceleration. The proton acceleration times for lepto-hadronic models are also much longer.

The GeV to radio response is a strong indication that synchrotron processes dominate the low energy emission component. A fast rise of the response profile after a delay of $\approx$43 days and a slow decay can be interpreted as an emitting region moving outwards in a conical jet: first becoming transparent to gamma rays, later to the radio. The fast rise after a delay may indicate a discontinuity in the jet properties.

\noindent\textit{Acknowledgements:} The important contributions from ETH Zurich grants ETH-10.08-2 and ETH-27.12-1 as well as the funding by the Swiss SNF and the German BMBF (Verbundforschung Astro- und Astroteilchenphysik) and HAP (Helmoltz Alliance for Astroparticle Physics) are gratefully acknowledged. Part of this work is supported by Deutsche Forschungsgemeinschaft (DFG) within the Collaborative Research Center SFB 876 "Providing Information by Resource-Constrained Analysis", project C3. We are thankful for the very valuable contributions from E. Lorenz, D. Renker and G. Viertel during the early phase of the project. We thank the Instituto de Astrof\'{\i}sica de Canarias for allowing us to operate the telescope at the Observatorio del Roque de los Muchachos in La Palma, the Max-Planck-Institut f\"ur Physik for providing us with the mount of the former HEGRA
CT3 telescope, and the MAGIC collaboration for their support. This research has made use of public data from the \textit{OVRO} 40-m telescope \citep{Richards_2011ApJS..194...29R}, the Bok Telescope on Kitt Peak and the 1.54 m Kuiper Telescope on Mt. Bigelow \citep{2009arXiv0912.3621S}, MAXI \citep{2009PASJ...61..999M}, \textit{Fermi}-LAT \citep{2009arXiv0912.3621S} and  \textit{Swift} \citep{2004NewAR..48..431G}.

\bibliographystyle{JHEP}
\setlength{\bibsep}{1pt}
\footnotesize{
\bibliography{main}

\providecommand{\href}[2]{#2}\begingroup\raggedright\begin{thebibliography}{10}

\bibitem{abdo_2011ApJ...736..131A}
A.~A. {Abdo}, M.~{Ackermann}, M.~{Ajello}, L.~{Baldini}, J.~{Ballet},
  G.~{Barbiellini} et~al., \emph{{Fermi Large Area Telescope Observations of
  Markarian 421: The Missing Piece of its Spectral Energy Distribution}},
  \href{https://doi.org/10.1088/0004-637X/736/2/131}{\emph{\apj} {\bfseries
  736} (2011) 131} [\href{https://arxiv.org/abs/1106.1348}{{\ttfamily
  1106.1348}}].

\bibitem{2015MNRAS.448..910C}
M.~{Cerruti}, A.~{Zech}, C.~{Boisson} and S.~{Inoue}, \emph{{A hadronic origin
  for ultra-high-frequency-peaked BL Lac objects}},
  \href{https://doi.org/10.1093/mnras/stu2691}{\emph{\mnras} {\bfseries 448}
  (2015) 910} [\href{https://arxiv.org/abs/1411.5968}{{\ttfamily 1411.5968}}].

\bibitem{2001APh....15..121M}
A.~{M{\"u}cke} and R.~J. {Protheroe}, \emph{{A proton synchrotron blazar model
  for flaring in Markarian 501}},
  \href{https://doi.org/10.1016/S0927-6505(00)00141-9}{\emph{Astroparticle
  Physics} {\bfseries 15} (2001) 121}
  [\href{https://arxiv.org/abs/astro-ph/0004052}{{\ttfamily
  astro-ph/0004052}}].

\bibitem{2014A&A...563A..91A}
J.~{Aleksi{\'c}}, L.~A. {Antonelli}, P.~{Antoranz}, A.~{Babic}, U.~{Barres de
  Almeida}, J.~A. {Barrio} et~al., \emph{{Rapid and multiband variability of
  the TeV bright active nucleus of the galaxy IC 310}},
  \href{https://doi.org/10.1051/0004-6361/201321938}{\emph{\aap} {\bfseries
  563} (2014) A91} [\href{https://arxiv.org/abs/1305.5147}{{\ttfamily
  1305.5147}}].

\bibitem{2004ApJ...601..759T}
C.~{Tanihata}, J.~{Kataoka}, T.~{Takahashi} and G.~M. {Madejski},
  \emph{{Evolution of the Synchrotron Spectrum in Markarian 421 during the 1998
  Campaign}}, \href{https://doi.org/10.1086/380779}{\emph{\apj} {\bfseries 601}
  (2004) 759} [\href{https://arxiv.org/abs/astro-ph/0310592}{{\ttfamily
  astro-ph/0310592}}].

\bibitem{2015A&A...576A.126A}
J.~{Aleksi{\'c}}, S.~{Ansoldi}, L.~A. {Antonelli}, P.~{Antoranz}, A.~{Babic},
  P.~{Bangale} et~al., \emph{{The 2009 multiwavelength campaign on Mrk 421:
  Variability and correlation studies}},
  \href{https://doi.org/10.1051/0004-6361/201424216}{\emph{\aap} {\bfseries
  576} (2015) A126} [\href{https://arxiv.org/abs/1502.02650}{{\ttfamily
  1502.02650}}].

\bibitem{2013arXiv1311.0478D}
D.~{Dorner}, A.~{Biland}, T.~{Bretz}, J.~{Buss}, S.~{Einecke}, D.~{Eisenacher}
  et~al., \emph{{FACT - Long-term Monitoring of Bright TeV-Blazars}},
  {\emph{arXiv e-prints} (2013) }
  [\href{https://arxiv.org/abs/1311.0478}{{\ttfamily 1311.0478}}].

\bibitem{2013JInst...8P6008A}
H.~{Anderhub}, M.~{Backes}, A.~{Biland}, V.~{Boccone}, I.~{Braun}, T.~{Bretz}
  et~al., \emph{{Design and operation of FACT - the first G-APD Cherenkov
  telescope}},
  \href{https://doi.org/10.1088/1748-0221/8/06/P06008}{\emph{Journal of
  Instrumentation} {\bfseries 8} (2013) P06008}
  [\href{https://arxiv.org/abs/1304.1710}{{\ttfamily 1304.1710}}].

\bibitem{2014JInst...9P0012B}
A.~{Biland}, T.~{Bretz}, J.~{Bu{\ss}}, V.~{Commichau}, L.~{Djambazov},
  D.~{Dorner} et~al., \emph{{Calibration and performance of the photon sensor
  response of FACT - the first G-APD Cherenkov telescope}},
  \href{https://doi.org/10.1088/1748-0221/9/10/P10012}{\emph{Journal of
  Instrumentation} {\bfseries 9} (2014) P10012}
  [\href{https://arxiv.org/abs/1403.5747}{{\ttfamily 1403.5747}}].

\bibitem{2013arXiv1307.6116K}
M.~L. {Knoetig}, A.~{Biland}, T.~{Bretz}, J.~{Bu{\ss}}, D.~{Dorner},
  S.~{Einecke} et~al., \emph{{FACT - Long-term stability and observations
  during strong Moon light}}, {\emph{arXiv e-prints} (2013) }
  [\href{https://arxiv.org/abs/1307.6116}{{\ttfamily 1307.6116}}].

\bibitem{2017ICRC...35..779H}
D.~{Hildebrand}, M.~L. {Ahnen}, M.~{Balbo}, A.~{Biland}, T.~{Bretz}, J.~{Buss}
  et~al., \emph{{Using Charged Cosmic Ray Particles to Monitor the Data Quality
  of FACT}}, {\emph{35th International Cosmic Ray Conference, Proceedings of
  Science} {\bfseries 301} (2017) 779}.

\bibitem{2017ICRC...35..612M}
M.~{Mahlke}, T.~{Bretz}, J.~{Adam}, L.~M. {Ahnen}, D.~{Baack}, M.~{Balbo}
  et~al., \emph{{FACT - Searching for periodicity in five-year light-curves of
  Active Galactic Nuclei}}, {\emph{35th International Cosmic Ray Conference,
  Proceedings of Science} {\bfseries 301} (2017) 612}.

\bibitem{2019arXiv190203875B}
T.~{Bretz}, \emph{{Zenith angle dependence of the cosmic ray rate as measured
  with imaging air-Cherenkov telescopes}},
  \href{https://doi.org/10.1016/j.astropartphys.2019.02.004}{\emph{Astroparticle
  Physics} {\bfseries 111} (2019) 72}
  [\href{https://arxiv.org/abs/1902.03875}{{\ttfamily 1902.03875}}].

\bibitem{2005ICRC....5..215R}
B.~{Riegel}, T.~{Bretz}, D.~{Dorner}, K.~{Berger} and D.~{H{\"o}hne}, \emph{{A
  systematic study of the interdependence of IACT image parameters}},
  {\emph{29th International Cosmic Ray Conference} {\bfseries 5} (2005) 215}.

\bibitem{2009ApJ...697.1071A}
W.~B. {Atwood}, A.~A. {Abdo}, M.~{Ackermann}, W.~{Althouse}, B.~{Anderson},
  M.~{Axelsson} et~al., \emph{{The Large Area Telescope on the Fermi Gamma-Ray
  Space Telescope Mission}},
  \href{https://doi.org/10.1088/0004-637X/697/2/1071}{\emph{\apj} {\bfseries
  697} (2009) 1071} [\href{https://arxiv.org/abs/0902.1089}{{\ttfamily
  0902.1089}}].

\bibitem{2013ApJS..207...19B}
W.~H. {Baumgartner}, J.~{Tueller}, C.~B. {Markwardt}, G.~K. {Skinner},
  S.~{Barthelmy}, R.~F. {Mushotzky} et~al., \emph{{The 70 Month Swift-BAT
  All-sky Hard X-Ray Survey}},
  \href{https://doi.org/10.1088/0067-0049/207/2/19}{\emph{\apjs} {\bfseries
  207} (2013) 19} [\href{https://arxiv.org/abs/1212.3336}{{\ttfamily
  1212.3336}}].

\bibitem{2005SSRv..120..165B}
D.~N. {Burrows}, J.~E. {Hill}, J.~A. {Nousek}, J.~A. {Kennea}, A.~{Wells},
  J.~P. {Osborne} et~al., \emph{{The Swift X-Ray Telescope}},
  \href{https://doi.org/10.1007/s11214-005-5097-2}{\emph{\ssr} {\bfseries 120}
  (2005) 165} [\href{https://arxiv.org/abs/astro-ph/0508071}{{\ttfamily
  astro-ph/0508071}}].

\bibitem{2009PASJ...61..999M}
M.~{Matsuoka}, K.~{Kawasaki}, S.~{Ueno}, H.~{Tomida}, M.~{Kohama}, M.~{Suzuki}
  et~al., \emph{{The MAXI Mission on the ISS: Science and Instruments for
  Monitoring All-Sky X-Ray Images}},
  \href{https://doi.org/10.1093/pasj/61.5.999}{\emph{\pasj} {\bfseries 61}
  (2009) 999} [\href{https://arxiv.org/abs/0906.0631}{{\ttfamily 0906.0631}}].

\bibitem{2009arXiv0912.3621S}
P.~S. {Smith}, E.~{Montiel}, S.~{Rightley}, J.~{Turner}, G.~D. {Schmidt} and
  B.~T. {Jannuzi}, \emph{{Coordinated Fermi/Optical Monitoring of Blazars and
  the Great 2009 September Gamma-ray Flare of 3C 454.3}}, {\emph{arXiv
  e-prints} (2009) } [\href{https://arxiv.org/abs/0912.3621}{{\ttfamily
  0912.3621}}].

\bibitem{roming_2005SSRv..120...95R}
P.~W.~A. {Roming}, T.~E. {Kennedy}, K.~O. {Mason}, J.~A. {Nousek}, L.~{Ahr},
  R.~E. {Bingham} et~al., \emph{{The Swift Ultra-Violet/Optical Telescope}},
  \href{https://doi.org/10.1007/s11214-005-5095-4}{\emph{\ssr} {\bfseries 120}
  (2005) 95} [\href{https://arxiv.org/abs/astro-ph/0507413}{{\ttfamily
  astro-ph/0507413}}].

\bibitem{Vaughan_2003MNRAS.345.1271V}
S.~{Vaughan}, R.~{Edelson}, R.~S. {Warwick} and P.~{Uttley}, \emph{{On
  characterizing the variability properties of X-ray light curves from active
  galaxies}},
  \href{https://doi.org/10.1046/j.1365-2966.2003.07042.x}{\emph{\mnras}
  {\bfseries 345} (2003) 1271}
  [\href{https://arxiv.org/abs/astro-ph/0307420}{{\ttfamily
  astro-ph/0307420}}].

\bibitem{Poutanen_2008MNRAS.389.1427P}
J.~{Poutanen}, A.~A. {Zdziarski} and A.~{Ibragimov}, \emph{{Superorbital
  variability of X-ray and radio emission of Cyg X-1 - II. Dependence of the
  orbital modulation and spectral hardness on the superorbital phase}},
  \href{https://doi.org/10.1111/j.1365-2966.2008.13666.x}{\emph{\mnras}
  {\bfseries 389} (2008) 1427}
  [\href{https://arxiv.org/abs/0802.1391}{{\ttfamily 0802.1391}}].

\bibitem{aleksic_2015A&A...578A..22A}
J.~{Aleksi{\'c}}, S.~{Ansoldi}, L.~A. {Antonelli}, P.~{Antoranz}, A.~{Babic},
  P.~{Bangale} et~al., \emph{{Unprecedented study of the broadband emission of
  Mrk 421 during flaring activity in March 2010}},
  \href{https://doi.org/10.1051/0004-6361/201424811}{\emph{\aap} {\bfseries
  578} (2015) A22} [\href{https://arxiv.org/abs/1412.3576}{{\ttfamily
  1412.3576}}].

\bibitem{Peterson_1998PASP..110..660P}
B.~M. {Peterson}, I.~{Wanders}, K.~{Horne}, S.~{Collier}, T.~{Alexander},
  S.~{Kaspi} et~al., \emph{{On Uncertainties in Cross-Correlation Lags and the
  Reality of Wavelength-dependent Continuum Lags in Active Galactic Nuclei}},
  \href{https://doi.org/10.1086/316177}{\emph{\pasp} {\bfseries 110} (1998)
  660} [\href{https://arxiv.org/abs/astro-ph/9802103}{{\ttfamily
  astro-ph/9802103}}].

\bibitem{2003ApJ...584L..53U}
P.~{Uttley}, R.~{Edelson}, I.~M. {McHardy}, B.~M. {Peterson} and
  A.~{Markowitz}, \emph{{Correlated Long-Term Optical and X-Ray Variations in
  NGC 5548}}, \href{https://doi.org/10.1086/373887}{\emph{\apjl} {\bfseries
  584} (2003) L53} [\href{https://arxiv.org/abs/astro-ph/0301216}{{\ttfamily
  astro-ph/0301216}}].

\bibitem{turler_1999A&A...349...45T}
M.~{T{\"u}rler}, T.~J.-L. {Courvoisier} and S.~{Paltani}, \emph{{Modelling the
  submillimetre-to-radio flaring behaviour of 3C 273}}, {\emph{\aap} {\bfseries
  349} (1999) 45} [\href{https://arxiv.org/abs/astro-ph/9906274}{{\ttfamily
  astro-ph/9906274}}].

\bibitem{esposito_2015A&A...576A.122E}
V.~{Esposito}, R.~{Walter}, P.~{Jean}, A.~{Tramacere}, M.~{T{\"u}rler},
  A.~{L{\"a}hteenm{\"a}ki} et~al., \emph{{The high energy spectrum of 3C 273}},
  \href{https://doi.org/10.1051/0004-6361/201424644}{\emph{\aap} {\bfseries
  576} (2015) A122} [\href{https://arxiv.org/abs/1503.02980}{{\ttfamily
  1503.02980}}].

\bibitem{Richards_2011ApJS..194...29R}
J.~L. {Richards}, W.~{Max-Moerbeck}, V.~{Pavlidou}, O.~G. {King}, T.~J.
  {Pearson}, A.~C.~S. {Readhead} et~al., \emph{{Blazars in the Fermi Era: The
  OVRO 40 m Telescope Monitoring Program}},
  \href{https://doi.org/10.1088/0067-0049/194/2/29}{\emph{\apjs} {\bfseries
  194} (2011) 29} [\href{https://arxiv.org/abs/1011.3111}{{\ttfamily
  1011.3111}}].

\bibitem{2004NewAR..48..431G}
N.~{Gehrels} and {Swift Team}, \emph{{The Swift {\ensuremath{\gamma}}-ray burst
  mission}}, \href{https://doi.org/10.1016/j.newar.2003.12.055}{\emph{New
  Astronomy Reviews} {\bfseries 48} (2004) 431}.

\end{thebibliography}\endgroup
}
\end{document}